\documentclass[aps,prl,twocolumn,superscriptaddress,showpacs]{revtex4}

\usepackage{graphicx}

\begin{document}
 
%-------------------------------------------------------------------------------

\title{The Quasielastic $^3$He$(e,e'p)d$ Reaction at Q$^2$~=~1.5~GeV$^2$ for 
Recoil Momenta up to~1~GeV/$c$}

\author{M.~M.~Rvachev}
\affiliation{Massachusetts Institute of Technology, Cambridge, Massachusetts 02139, USA}
\author{F.~Benmokhtar}
\affiliation{Rutgers, The State University of New Jersey, Piscataway, New Jersey 08854, USA}
\affiliation{Universit\'e des Sciences et de la Technologie, BP 32, El Alia, Bab Ezzouar, 16111 Alger, Alg\'erie}
\author{E.~Penel-Nottaris}
\affiliation{Laboratoire de Physique Subatomique et de Cosmologie, F-38026 Grenoble, France}
\author{K.~A.~Aniol}
\affiliation{California State University Los Angeles, Los Angeles, California 90032, USA}
\author{W.~Bertozzi}
\affiliation{Massachusetts Institute of Technology, Cambridge, Massachusetts 02139, USA}
\author{W.~U.~Boeglin}
\affiliation{Florida International University, Miami, Florida 33199, USA}
\author{F.~Butaru}
\affiliation{Laboratoire de Physique Subatomique et de Cosmologie, F-38026 Grenoble, France}
\author{J.~R.~Calarco}
\affiliation{University of New Hampshire, Durham, New Hampshire 03824, USA}
\author{Z.~Chai}
\affiliation{Massachusetts Institute of Technology, Cambridge, Massachusetts 02139, USA}
\author{C.~C.~Chang}
\affiliation{University of Maryland, College Park, Maryland 20742, USA}
\author{J.~-P.~Chen}
\affiliation{Thomas Jefferson National Accelerator Facility, Newport News, Virginia 23606, USA}
\author{E.~Chudakov}
\affiliation{Thomas Jefferson National Accelerator Facility, Newport News, Virginia 23606, USA}
\author{E.~Cisbani}
\affiliation{INFN, Sezione Sanit\'a and Istituto Superiore di Sanit\'a, Laboratorio di Fisica, I-00161 Rome, Italy}
\author{A.~Cochran}
\affiliation{Hampton University, Hampton, Virginia 23668, USA}
\author{J.~Cornejo}
\affiliation{California State University Los Angeles, Los Angeles, California 90032, USA}
\author{S.~Dieterich}
\affiliation{Rutgers, The State University of New Jersey, Piscataway, New Jersey 08854, USA}
\author{P.~Djawotho}
\affiliation{College of William and Mary, Williamsburg, Virginia 23187, USA}
\author{W.~Duran}
\affiliation{California State University Los Angeles, Los Angeles, California 90032, USA}
\author{M.~B.~Epstein}
\affiliation{California State University Los Angeles, Los Angeles, California 90032, USA}
\author{J.~M.~Finn}
\affiliation{College of William and Mary, Williamsburg, Virginia 23187, USA}
\author{K.~G.~Fissum}
\affiliation{University of Lund, Box 118, SE-221 00 Lund, Sweden} 
\author{A.~Frahi-Amroun}
\affiliation{Universit\'e des Sciences et de la Technologie, BP 32, El Alia, Bab Ezzouar, 16111 Alger, Alg\'erie}
\author{S.~Frullani}
\affiliation{INFN, Sezione Sanit\'a and Istituto Superiore di Sanit\'a, Laboratorio di Fisica, I-00161 Rome, Italy}
\author{C.~Furget}
\affiliation{Laboratoire de Physique Subatomique et de Cosmologie, F-38026 Grenoble, France}
\author{F.~Garibaldi}
\affiliation{INFN, Sezione Sanit\'a and Istituto Superiore di Sanit\'a, Laboratorio di Fisica, I-00161 Rome, Italy}
\author{O.~Gayou}
\affiliation{College of William and Mary, Williamsburg, Virginia 23187, USA}
\author{S.~Gilad}
\affiliation{Massachusetts Institute of Technology, Cambridge, Massachusetts 02139, USA}
\author{R.~Gilman}
\affiliation{Rutgers, The State University of New Jersey, Piscataway, New Jersey 08854, USA}
\affiliation{Thomas Jefferson National Accelerator Facility, Newport News, Virginia 23606, USA}
\author{C.~Glashausser}
\affiliation{Rutgers, The State University of New Jersey, Piscataway, New Jersey 08854, USA}
\author{J.-O.~Hansen}
\affiliation{Thomas Jefferson National Accelerator Facility, Newport News, Virginia 23606, USA}
\author{D.~W.~Higinbotham}
\affiliation{Massachusetts Institute of Technology, Cambridge, Massachusetts 02139, USA}
\affiliation{Thomas Jefferson National Accelerator Facility, Newport News, Virginia 23606, USA}
\author{A.~Hotta}
\affiliation{University of Massachusetts, Amherst, Massachusetts 01003, USA}
\author{B.~Hu}
\affiliation{Hampton University, Hampton, Virginia 23668, USA}
\author{M.~Iodice}
\affiliation{INFN, Sezione Sanit\'a and Istituto Superiore di Sanit\'a, Laboratorio di Fisica, I-00161 Rome, Italy}
\author{R.~Iomni}
\affiliation{INFN, Sezione Sanit\'a and Istituto Superiore di Sanit\'a, Laboratorio di Fisica, I-00161 Rome, Italy}
\author{C.~W.~de~Jager}
\affiliation{Thomas Jefferson National Accelerator Facility, Newport News, Virginia 23606, USA}
\author{X.~Jiang}
\affiliation{Rutgers, The State University of New Jersey, Piscataway, New Jersey 08854, USA}
\author{M.~K.~Jones}
\affiliation{Thomas Jefferson National Accelerator Facility, Newport News, Virginia 23606, USA}
\affiliation{University of Maryland, College Park, Maryland 20742, USA}
\author{J.~J.~Kelly}
\affiliation{University of Maryland, College Park, Maryland 20742, USA}
\author{S.~Kox}
\affiliation{Laboratoire de Physique Subatomique et de Cosmologie, F-38026 Grenoble, France}
\author{M.~Kuss}
\affiliation{Thomas Jefferson National Accelerator Facility, Newport News, Virginia 23606, USA}
\author{J.~M.~Laget}
\affiliation{CEA-Saclay, F-91191 Gif –Sur-Yvette Cedex, France}
\author{R.~De~Leo}
\affiliation{INFN, Sezione di Bari and University of Bari, I-70126 Bari, Italy}
\author{J.~J.~LeRose}
\affiliation{Thomas Jefferson National Accelerator Facility, Newport News, Virginia 23606, USA}
\author{E.~Liatard}
\affiliation{Laboratoire de Physique Subatomique et de Cosmologie, F-38026 Grenoble, France}
\author{R.~Lindgren}
\affiliation{University of Virginia, Charlottesville, Virginia 22901, USA}
\author{N.~Liyanage}
\affiliation{Thomas Jefferson National Accelerator Facility, Newport News, Virginia 23606, USA}
\author{R.~W.~Lourie}
\affiliation{State University of New York at Stony Brook, Stony Brook, New York 11794, USA}
\author{S.~Malov}
\affiliation{Rutgers, The State University of New Jersey, Piscataway, New Jersey 08854, USA}
\author{D.~J.~Margaziotis}
\affiliation{California State University Los Angeles, Los Angeles, California 90032, USA}
\author{P.~Markowitz}
\affiliation{Florida International University, Miami, Florida 33199, USA}
\author{F.~Merchez}
\affiliation{Laboratoire de Physique Subatomique et de Cosmologie, F-38026 Grenoble, France}
\author{R.~Michaels}
\affiliation{Thomas Jefferson National Accelerator Facility, Newport News, Virginia 23606, USA}
\author{J.~Mitchell}
\affiliation{Thomas Jefferson National Accelerator Facility, Newport News, Virginia 23606, USA}
\author{J.~Mougey}
\affiliation{Laboratoire de Physique Subatomique et de Cosmologie, F-38026 Grenoble, France}
\author{C.~F.~Perdrisat}
\affiliation{College of William and Mary, Williamsburg, Virginia 23187, USA}
\author{V.~A.~Punjabi}
\affiliation{Norfolk State University, Norfolk, Virginia 23504, USA}
\author{G.~Qu\'em\'ener}
\affiliation{Laboratoire de Physique Subatomique et de Cosmologie, F-38026 Grenoble, France}
\author{R.~D.~Ransome}
\affiliation{Rutgers, The State University of New Jersey, Piscataway, New Jersey 08854, USA}
\author{J.-S.~R\'eal}
\affiliation{Laboratoire de Physique Subatomique et de Cosmologie, F-38026 Grenoble, France}
\author{R.~Roch\'e}
\affiliation{Florida State University, Tallahassee, Florida 32306, USA}
\author{F.~Sabati\'e}
\affiliation{Old Dominion University, Norfolk, Virginia 23529, USA}
\author{A.~Saha}
\affiliation{Thomas Jefferson National Accelerator Facility, Newport News, Virginia 23606, USA}
\author{D.~Simon}
\affiliation{Old Dominion University, Norfolk, Virginia 23529, USA}
\author{S.~Strauch}
\affiliation{Rutgers, The State University of New Jersey, Piscataway, New Jersey 08854, USA}
\author{R.~Suleiman}
\affiliation{Massachusetts Institute of Technology, Cambridge, Massachusetts 02139, USA}
\author{T.~Tamae}
\affiliation{Tohoku University, Sendai 980, Japan} 
\author{J.~A.~Templon}
\affiliation{University of Georgia, Athens, Georgia 30602, USA}
\author{R.~Tieulent}
\affiliation{Laboratoire de Physique Subatomique et de Cosmologie, F-38026 Grenoble, France}
\author{H.~Ueno}
\affiliation{Yamagata University, Kojirakawa-machi 1-4-12, Yamagata 990-8560, Japan}
\author{P.~E.~Ulmer}
\affiliation{Old Dominion University, Norfolk, Virginia 23529, USA}
\author{G.~M.~Urciuoli}
\affiliation{INFN, Sezione Sanit\'a and Istituto Superiore di Sanit\'a, Laboratorio di Fisica, I-00161 Rome, Italy}
\author{E.~Voutier}
\affiliation{Laboratoire de Physique Subatomique et de Cosmologie, F-38026 Grenoble, France}
\author{K.~Wijesooriya}
\affiliation{University of Illinois at Urbana Champaign, Urbana, Illinois 61801, USA}
\author{B.~Wojtsekhowski}
\affiliation{Thomas Jefferson National Accelerator Facility, Newport News, Virginia 23606, USA}

\collaboration{The Jefferson Lab Hall A Collaboration}

\date{\today}

\begin{abstract}
We have studied the quasielastic $^3$He$(e,e'p)d$ reaction in perpendicular 
coplanar kinematics, with the energy and momentum transferred by the electron 
fixed at 840 MeV and 1502 MeV/$c$, respectively. The $^3$He$(e,e'p)d$ cross 
section was measured for missing momenta up to 1000 MeV/$c$, while the $A_{TL}$ 
asymmetry was extracted for missing momenta up to 660 MeV/$c$. For missing 
momenta up to 150 MeV/$c$, the measured cross section is described well by 
calculations that use a variational ground-state wave function of the $^3$He 
nucleus derived from a potential that includes three-body forces. For missing 
momenta from 150 to 750 MeV/$c$, strong final-state interaction effects are 
observed. Near 1000~MeV/$c$, the experimental cross section is more than an 
order of magnitude larger than predicted by available theories. The $A_{TL}$ 
asymmetry displays characteristic features of broken factorization, and is 
described reasonably well by available models.

\end{abstract}

\pacs{21.45.+v, 25.30.Dh, 27.10+h}
%\keywords{}

\maketitle

%-------------------------------------------------------------------------------
%
% Introduction
%
%-------------------------------------------------------------------------------

Microscopic calculations make it possible now to calculate the bound-state and scattering-state wave functions from Hamiltonian models for processes involving three-nucleon systems~\cite{marcucci}.  Thus, using modern (non-relativistic) Faddeev~\cite{Gloeckle00,Gloeckle02} and variational~\cite{schi} techniques to solve the three-body problem, one hopes to test the ability to predict the structure of three-body systems with state-of-the-art realistic NN potentials. The quasielastic $^3$He$(e,e'p)d$ reaction has been used to study the single-proton wave function in $^3$He. In the Plane-Wave Impulse Approximation (PWIA), this reaction samples the single-particle momentum distribution in the $^3$He nucleus. However, reaction-dynamics processes such as final-state interactions (FSI), two-body currents (meson exchange and isobar), as well as relativity have to be taken into account in the data interpretation. Unfortunately, the above mentioned modern computational techniques are not yet suffiently developed to reliably describe the reaction dynamics at high energies.  As such they would tremendously benefit from data for guidance in their development process.  

High-energy electron beams with high currents and 100\% duty factor at the 
Thomas Jefferson National Accelerator Facility (JLab) enable experiments to 
reach new kinematic domains and levels of precision in utilizing the $(e,e'p)$ 
reaction to study nuclear structure and reaction dynamics.  In this Letter, we address the interplay between nuclear structure and reaction dynamics by 
%reporting on precision measurements~\cite{E89044} 
providing an extensive and precise data set that includes cross sections and the $A_{TL}$ asymmetry for the $^3$He$(e,e'p)d$ reaction in constant quasielastic electron kinematics.  This data set significantly extends the available data in both the transferred four-momentum and the recoil momentum of the undetected deuteron (missing momentum), $p_m$.

%-------------------------------------------------------------------------------
%
% Experiment description
%
%-------------------------------------------------------------------------------

Measurements were performed using an incident beam of 4806 MeV and the two 
high-resolution spectrometer system (HRS) in Hall A of JLab. A detailed 
description of the Hall A instrumentation is available in~\cite{NIM}. Electrons 
(protons) were detected with the left (right) HRS respectively, HRS-L and 
HRS-R. Scattered electrons were detected at a central scattering angle of 
16.4$^\circ$ and a central momentum of 3966 MeV/$c$, corresponding to the 
quasielastic knockout of protons from the $^3$He nucleus with transferred 
three-momentum $|\vec{q} \, | = 1502$ MeV/$c$, energy $\omega = 840$~MeV, 
four-momentum $Q^2$ = 1.55~GeV$^2$, and Bjorken scaling variable 
$x_B=Q^2/(2\omega M_p) = 0.98$. The range in accepted $Q^2$ and $\omega$ was $\pm$ 0.12 GeV$^2$ 
and $\pm$ 20 MeV respectively.  The ejected proton was detected in coincidence 
with the scattered electron in coplanar kinematics over a range of angles and 
momenta (see Table 1), to measure the $p_m$ dependence of the 
$^3$He$(e,e'p)d$ cross section on both sides of the momentum-transfer direction.

\begin{table}[ht]
\caption{Central kinematic values of the $^3$He$(e,e'p)d$ measurements. 
Listed are central settings of the hadron (HRS-R) spectrometer (momentum $P_p$, angle 
$\theta_p$, and missing momentum $p_m$). Negative (positive) $p_m$ corresponds 
to the detected proton forward (backward) of $\vec{q}$. The electron 
kinematics were fixed, at incident and scattered electron energies of 
E=4806 MeV and E$'$=3966 MeV, respectively, and scattering angle of 
$\theta_e$=16.4$^\circ$ ($Q^2$=1.55~GeV$^2$, $|\vec{q} \, |$=1502~MeV/$c$, 
$\omega$=840~MeV, $x_B$=0.98). }
\begin{tabular}{|c|c|c|}\hline
\hspace{9mm}$p_m$\hspace{9mm} &\hspace{9mm} $P_p$\hspace{9mm}   & \hspace{9mm} 
$\theta_p$ \hspace{9mm} \\
MeV/$c$ & MeV/$c$ & deg \\ \hline
-550   &   1406   & 26.79    \\
-425   &   1444   & 31.84    \\
-300   &   1472   & 36.76    \\
-150   &   1493   & 42.56    \\
0   &   1500   & 48.30    \\
150   &   1493   & 54.04    \\
300   &   1472   & 59.83    \\
425   &   1444   & 64.76    \\
550   &   1406   & 69.80    \\
750   &   1327   & 78.28    \\
1000   &   1171   & 89.95    \\
%$p_m$ & $P_p$   & $\theta_p$ & $\epsilon_{syst}$ \\
%MeV/$c$ & MeV/$c$ & deg & \% \\\hline
%-550   &   1406   & 26.79  &   4.8    \\
%-425   &   1444   & 31.84  &   5.2    \\
%-300   &   1472   & 36.76  &   5.1    \\
%-150   &   1493   & 42.56  &   9.0    \\
%0   &   1500   & 48.30  &   5.4    \\
%150   &   1493   & 54.04  &   7.2    \\
%300   &   1472   & 59.83  &   5.4    \\
%425   &   1444   & 64.76  &   4.6    \\
%550   &   1406   & 69.80  &   5.7    \\
%750   &   1327   & 78.28  &   4.6    \\
%1000   &   1171   & 89.95  &   4.6    \\
\hline

\end{tabular}

\end{table}

A cooled, 10.3~cm-diameter $^3$He gas target was used at temperature 
T~=~6.3~K and pressures P = 8.30 - 10.9~atm, corresponding to densities 
$\rho$ = 0.0603 - 0.0724~g/cm$^3$.  Relative changes in the target density were 
monitored by observing changes in the rate of singles events in the fixed 
HRS-L per unit beam charge passing through the target. The target density was 
determined by measuring the elastic $^3$He$(e,e)$ cross section at a beam 
energy of 644 MeV ($\theta_e=30.7^\circ$, $Q^2$ = 0.11 GeV$^2$, and 
normalizing it to the cross section derived from a fit to the world data of 
$^3$He elastic form factors~\cite{amro94}.  The overall normalization 
uncertainty of the $^3$He density is estimated to be 2.9\%, obtained as the 
quadratic sum of the systematic uncertainty of our $^3$He elastic cross section 
measurement (2.4\%), the statistical uncertainty (0.5\%), the uncertainty in 
the $^3$He form factors (1.5\%), and a 0.5\% uncertainty due to possible 
fluctuations in the target density during the change of the beam energy from 
4806 to 644 MeV.  

Event triggers were formed by coincident signals from scintillator arrays. 
Particle tracks were reconstructed using the HRS vertical drift chambers. 
The small $\pi^-$ background in the HRS-L was rejected using a CO$_2$ Gas 
\v{C}erenkov detector.  In the HRS-R, coincident $\pi^+$, $^2$H, and $^3$H 
were separated from the protons using the time difference between particles 
detected in the two spectrometers.  Most of the accidental coincident events 
were rejected by cuts on the difference between interaction points in the 
target along the beam as reconstructed by the two spectrometers, $|z_h - z_e| \leq$ 2~cm, where the interaction-point resolution was about 8~mm (FWHM), and on the $E_m$ (missing energy) spectrum. The $E_m$ resolution was 2.4 MeV (FWHM).  The remaining accidental background was 
subtracted using the coincidence timing between the spectrometers.  Events 
originating in the target Al walls were rejected by requiring reconstructed 
events to originate within 3.5 cm from the target center.  With these cuts, the signal/noise ratio in the most extreme kinematics, for $p_m \approx 1$ GeV/$c$, was 50/1 and in the worse case, for $p_m$ $\approx$ -600 -- -700 MeV/$c$, about 0.8/1 - see Fig.~1.

In the cross-section analysis, a flat acceptance region of both HRSs was 
defined using an R-function cut imposed on the target variables.  An 
R-function is a function whose sign is completely determined by the signs of 
its arguments \cite{rvac95,rvac82}.  Using constructive-geometrical properties 
of R-functions, one can define a complicated multidimensional acceptance 
region as an analytical expression, and vary the region's boundaries until the 
phase space is maximized within the flat acceptance region of the 
spectrometers~\cite{rvac03}. The use of R-functions allowed us to double the 
accepted 
phase space compared to the commonly used rectangular cuts on target variables.

%-------------------------------------------------------------------------------
%
% Cross section data
%
%-------------------------------------------------------------------------------

\begin{figure}[ht]
\includegraphics[width=3.25in]{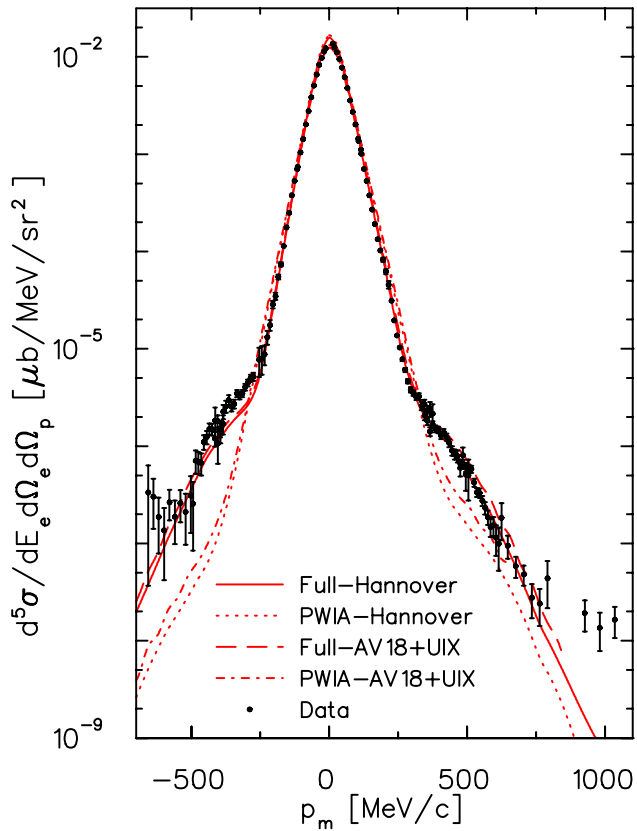}
\caption{Measured $^3$He$(e,e'p)d$ cross section as a function of the missing 
momentum, $p_m$.  Also displayed are PWIA and full 
calculations in the diagrammatic approach for two different ground-state wave function. 
}
\label{fig:xs}
\end{figure}

The $^3$He$(e,e'p)d$ cross section was extracted using the simulation program 
MCEEP \cite{ulme00} taking into account the effects of internal and external 
radiation, particle energy loss, deviations from monochromaticity of the 
beam, and spectrometer resolutions. 
For each $p_m$ bin, the simulated yields were varied by modifying the 
spectral function used in MCEEP to achieve calculated cross sections that agreed
with the measured ones in both the $^3$He$(e,e'p)d$ $E_m$ bin 
and the adjacent $^3$He$(e,e'p)pn$ $E_m$ bin~\cite{rvac03}.
Cross sections were extracted from the re-weighted $^3$He$(e,e'p)d$ yield, 
corrected for radiation, and for contributions from $^3$He$(e,e'p)pn$ to each 
$^3$He$(e,e'p)d$ kinematic bin.  On average, these contributions were about 3\%. 
Within each bin, the simulated $^3$He$(e,e'p)$ cross section was assumed to 
depend on the $\sigma_{cc1}$ prescription of de~Forest \cite{cc1} for the 
off-shell electron-proton cross section. This technique allows one to separate 
the $p_m$ dependence of the reaction from the rapid dependence on the electron 
kinematics~\cite{rvac03}. In addition to the over-all normalization uncertainty 
(2.9\%, see above), the over-all systematic uncertainty was 3.4\% dominated by 
uncertainties in the solid angle (2.0\%), the selection ($E_m$ cut) of the 
two-body break-up reaction channel (1.5\%) and the knowledge of the effective 
target length via a cut on the interaction vertex location (1.4\%). 

The extracted $^3$He$(e,e'p)d$ cross section is plotted in Fig.~1 as a function 
of $p_m$. We note that the range of $p_m$ measured (resulting in measured 
cross-section values varying over six orders of magnitude), is significantly 
larger than in any other previous measurement. Moreover, contrary to previous 
experiments~\cite{He3exp,He3exp2,He3exp3}, our measurements over this entire range were performed at fixed electron kinematics.

Also displayed in Fig.~1 are four theoretical curves by Laget.  The PWIA and full Hannover calculations use the Hannover bound-nucleon wave function~\cite{sauer} 
corresponding to the solution to the three-body Faddeev equation with the 
Paris NN potential and no three-body forces.  The AV18+UIX curves are the 
same PWIA and full calculations respectively, but with a bound-state 
nuclear wave function derived by a variational technique using the 
Argonne V18 NN potential and the Urbana IX three-body force~\cite{jforest}.  
All calculations use a diagrammatic approach. The kinematics as well as the 
nucleon and meson propagators are relativistic, and no restricted angular 
(Glauber type) approximation has been made in the various loop integrals.  
Details of the model can be found in~\cite{lage}. The PWIA curves 
include only one-body interactions, while the full calculations include FSI, 
meson ($\pi$ and $\rho$) exchange and intermediate $\Delta$ formation currents 
as well as three-body (three nucleon $\pi$ double scattering) amplitudes.  
The FSI in these calculations follow a global parameterization of the NN 
scattering amplitude, obtained from experiments in LANL, SATURNE and 
COSY~\cite{lage03}. On the scale of Fig.~1, the differences between the calculations 
using the two ground-state wave functions are very small. By far, FSI 
constitute the major difference between the full and PWIA  
calculations. Meson exchange and intermediate $\Delta$ current 
contributions are generally small (up to 20-25\%), and the three-body 
contributions are negligible~\cite{lage03}. 

Three regions of $p_m$ can be discerned in Fig.~1. For $|{\vec {p}}_m|$ 
below $\sim$ 150 MeV/$c$, roughly within the Fermi momentum, the deuteron can be viewed as only marginally involved in the interaction. Hence, the data are expected to be dominated 
by the single-proton characteristics of the $^3$He wave function. As can be 
observed, both the PWIA and full curves describe the data quite well, and the 
difference between them is rather small - see also Fig.~2 for details. For $|{\vec {p}}_m|$ between 150 and 750 MeV/$c$, well above the Fermi momentum, the cross section is expected to be dominated by the dynamics of the reaction.  Indeed, very large contributions 
from dynamical effects are observed. While 
%Laget's 
the full calculations describe 
the data very well, the PWIA curves over-predict the data by up to a factor of 2 for $150 \leq |{\vec {p}}_m| \leq$ 300 MeV/$c$ and under-predict them by up to an order of magnitude for $300 \leq |{\vec {p}}_m| \leq 750$ MeV/$c$.
The differences between the two PWIA and two full-calculation curves 
%by Laget 
are very much dominated by 
FSI. At $x_B$=1, the on-shell rescattering of the fast nucleon on a nucleon 
at rest is preferred and the contribution of FSI is maximal. Because the NN 
scattering amplitude is almost purely absorptive in the JLab energy range, the 
corresponding FSI amplitude interferes destructively with the PWIA amplitude 
below, and constructively above $p_m \approx$ 300~MeV/$c$~\cite{lage03}. We note the difference in cross sections in this region for negative and positive $p_m$, and it is discussed below.  For $p_m$ larger than 750 MeV/$c$, 
%Laget's 
the calculations gradually deviate from the experimental 
data: at 1000~MeV/$c$, they grossly under-predict the measured cross section by 
more than an order of magnitude. Whether it is a consequence of the truncation of the diagrammatic expansion or a signature of other degrees of freedom is an open question.
 
%-------------------------------------------------------------------------------
%
% Wave functions comparison
%
%-------------------------------------------------------------------------------

\begin{figure}[ht]
\includegraphics[angle=90,width=3.25in]{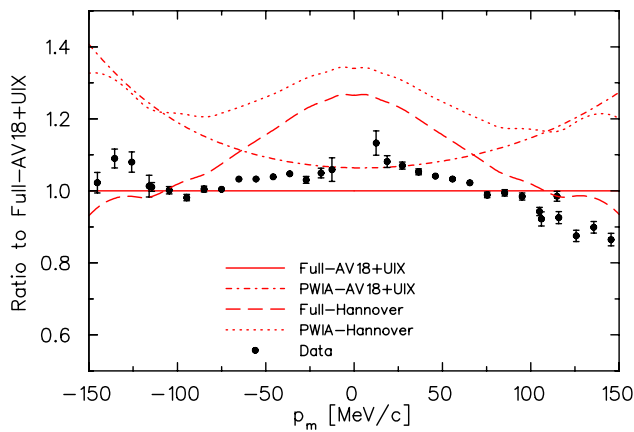}
%\caption{Same data as in Fig.~1 for low $p_m$ only, but shown 
%as a ratio to Laget's full calculations using the Hannover 
%ground-state wave function (gswf). Also shown are the ratios to the full 
%calculations that use the Hannover gswf of Laget's full calculations using the 
%gswf generated from the AV18 NN potential and the Urbana IX three-nucleon 
%force, as well as the two corresponding PWIA curves. See text for details.}
%
\caption{Same data as in Fig.~1 for low $p_m$ only, but shown
as a ratio to 
%Laget's 
the full calculations using the ground-state wave function (gswf) 
generated from the AV18 NN potential and the Urbana IX three-nucleon
force.
Also shown are the ratios to this calculation of 
%Laget's 
the full calculations that use the Hannover gswf, as well as of the two corresponding PWIA curves.}
\label{fig:spf}
\end{figure}

The sensitivity of the data to the details of the wave function at low 
$|{\vec {p}}_m|$ is shown in Fig.~2. In order to enhance the details, Fig.~2 
displays the low $|{\vec {p}}_m|$ subset of the data from Fig.~1 as a ratio to 
%Laget's 
the full calculations using the AV18 NN potential and the Urbana IX three-nucleon force.
Also displayed are the ratios to the same calculation of 
%Laget's 
the full Hannover ground-state wave function and the two corresponding PWIA curves. As already noted, in 
the low $|{\vec {p}}_m|$ region, we expect reaction effects such as FSI and 
two-body currents to be relatively small as compared to higher $|{\vec {p}}_m|$ 
(Fig.~\ref{fig:xs}), and hence the data to be more sensitive to the details of the calculated ground-state wave functions than to the uncertainty in describing reaction dynamics.  As can be seen in the figure, the curves produced by this model are mainly sensitive to 
the details of the bound-nucleon wave function. We note that for $p_m$ below 50~MeV/$c$, the 
calculations are purely co-planar perpendicular kinematics whereas 
experimentally, because of the large $|\vec {q}|$, it is difficult to avoid 
contaminations with parallel and out-of-plane components. For 
$|{\vec {p}}_m|> \, $50 MeV/$c$, we observe that the curve that best agree with the data is the full AV18+UIX. We suggest that this better agreement with 
the data is related to the fact that the wave function generated from the 
AV18+UIX potentials reproduces the correct 
$^3$He binding energy (by construction), while the Hannover wave function that does not include three-body forces underbinds the $^3$He by $\sim$ 0.7 MeV.

%-------------------------------------------------------------------------------
%
% A_TL results
%
%-------------------------------------------------------------------------------

The $A_{TL}$ asymmetry was extracted for $0 \leq |{\vec {p}}_m| \leq 
660$~MeV/$c$ according to 
\begin{equation}
A_{TL}=\frac{\sigma_{+}-\sigma_{-}}{\sigma_{+}+\sigma_{-}},
\end{equation}
where $\sigma_{+}$ and $\sigma_{-}$ are coplanar $^3$He$(e,e'p)d$ cross 
sections measured at positive and negative missing momentum respectively.
%right and left of the transferred-momentum $\vec{q}$ direction. 
%In the extraction of $A_{TL}$, the phase-space acceptances of kinematics bins %on both sides of $\vec{q}$ were matched in $p_m$, $\omega$, 
%and $|\vec{q} \, |$. 
The $A_{TL}$ observable downplays the significance of the ground-state wave 
function, by virtue of the ratio involved in its definition~\cite{kell96} and 
there exist indications that it is sensitive to relativistic 
effects~\cite{gilad} and to mechanisms that break the simple factorization 
scheme of PWIA cross sections~\cite{udiap}.  

\begin{figure}[htb]
\includegraphics[width=3.25in]{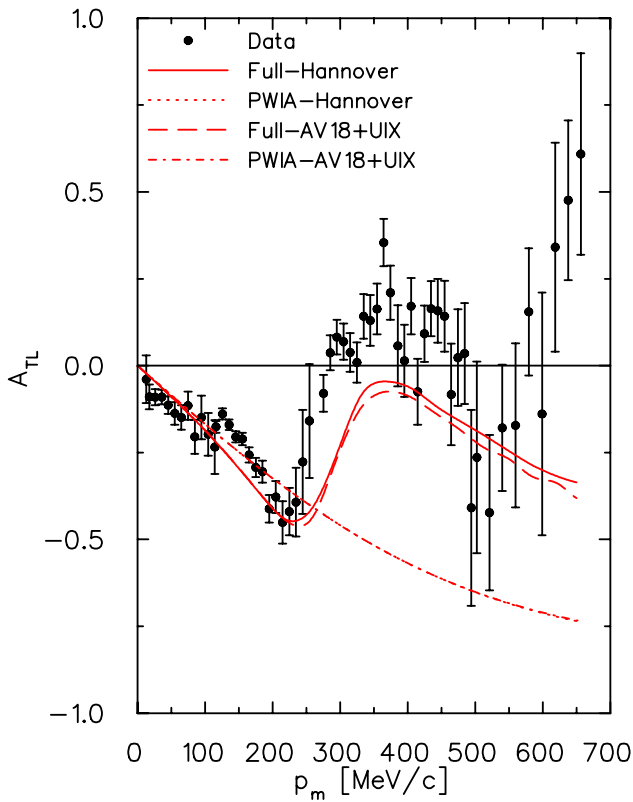}
\caption{The measured $A_{TL}$ asymmetry.  The curves are the 
same four calculations 
%by Laget 
used in Figs.~1 and 2; by definition, the two 
PWIA curves are indistinguishable.}
\label{fig:alt}
\end{figure}

Figure~\ref{fig:alt} displays the extracted $A_{TL}$ data with the PWIA and 
full calculations 
%by Laget 
using the two ground-state wave functions described 
above. The difference in the two ground-state wave functions has a very small 
effect in the full calculations. In contrast to the PWIA calculations, the 
measured $A_{TL}$ displays a structure characteristic of broken 
factorization~\cite{udiap}: the oscillating pattern of $A_{TL}$ comes directly 
from the interference between different reaction amplitudes. Both 
%of Laget's 
full calculations describe the data reasonably well by displaying 
similar structure. Such structure in $A_{TL}$ was previously observed in the 
quasielastic removal of p-shell protons in the $^{16}$O$(e,e'p)$ 
reaction~\cite{gao}, and was well reproduced by relativistic Distorted-Wave 
Impulse Approximation calculations by Udias {\it{et al.}}~\cite{udia99}. In 
that case broken factorization was attributed to dynamical relativistic 
effects, enhancement of the lower components of the Dirac spinors. However, these effects are marginal in our experiment because of the low 
nuclear density of $^3$He \cite{udiasrila}. Rather, in our case the 
factorization is broken by the strong interference between the PWIA 
and re-scattering amplitudes~\cite{lage03}.

%-------------------------------------------------------------------------------
%
% Conclusions
%
%-------------------------------------------------------------------------------

In summary, we measured the $^3$He$(e,e'p)d$ cross sections and $A_{TL}$ asymmetry 
at $Q^2 = 1.55$~GeV$^2$ and $x_B = 0.98$. For $|{\vec {p}}_m|$ below 150 MeV/$c$ 
the data are mostly sensitive to the details of the bound-state wave function. 
The best agreement is observed with calculations using a $^3$He ground-state wave 
function generated from the Argonne V18 NN potential and the Urbana IX 
three-nucleon force, which also better reproduces the $^3$He binding energy. For 
$|{\vec {p}}_m|$ from 150 to 750 MeV/$c$, strong FSI effects are observed as 
quenching (enhancement) of the cross section below (above) $|{\vec {p}}_m|$ of 
about 300~MeV/$c$. For missing momenta from 750 to 1000~MeV/$c$, the measured 
$^3$He$(e,e'p)d$ cross sections are increasingly larger (more than an order of 
magnitude at 1000~MeV/$c$) than predicted by available theories. Whether it is a 
consequence of the truncation of the diagrammatic expansion or a signature of the 
existence of exotic effects is an open question. The measured $A_{TL}$ displays 
strong structure characteristic of broken factorization due to interference 
between the PWIA and re-scattering amplitudes. Calculations 
%by Laget 
using a diagrammatic method well 
describe all observables up to $|{\vec {p}}_m|$ = 750 MeV/$c$. Other 
calculations of this reaction~\cite{{udiasrila},{ciofi},{schia}} that have recently become available similarly interpret the data.

%-------------------------------------------------------------------------------
%
% Acknowledgements
%
%-------------------------------------------------------------------------------

This work was supported by the U.S. Department of Energy (DOE) 
contract DE-AC05-84ER40150 Modification No. M175, under which the Southern Universities Research 
Association (SURA) operates the Thomas Jefferson National Accelerator 
Facility, DOE contract DE-FC02-94ER40818, other DOE contracts, the National 
Science Foundation, the Italian Istituto Nazionale di Fisica Nucleare (INFN), 
the French Atomic Energy Commission and National Center of Scientific Research, 
the Natural Science and Engineering Research Council of Canada, and Grant-in-Aid 
for Scientific Research (KAKENHI) (No.\ 14540239) from the Japan Society for 
Promotion of Science (JSPS).

\bibliography{Marat-prl}

\end{document}